# Theoretical investigation of delafossite-$Cu_2ZnSnO_4$ as a promising photovoltaic absorber

Seoung-Hun Kang[1,a)], Myeongjun Kang[2,a)], Sang Woon Hwang[3], Sinchul Yeom[1], Mina Yoon[1], Jong Mok Ok[2,b)], Sangmoon Yoon[3,b)]

[1] Materials Science and Technology Division, Oak Ridge National Laboratory, Oak Ridge, Tennessee 37831, USA

[2] Department of Physics, Pusan National University, Busan 46241, Republic of Korea

[3] Department of Physics, Gachon University, Seongnam 13120, Republic of Korea

In the quest for efficient and cost-effective photovoltaic absorber materials beyond silicon, considerable attention has been directed toward exploring alternatives. One such material, zincblende-derived $Cu_2ZnSnS_4$ (CZTS), has shown promise due to its ideal band-gap size and high absorption coefficient. However, challenges such as structural defects and secondary phase formation have hindered its development. In this study, we examine the potential of another compound $Cu_2ZnSnO_4$ (CZTO) with a similar composition to CZTS as a promising alternative. Employing *ab initio* density function theory (DFT) calculations in combination with an evolutionary structure prediction algorithm, we identify that the crystalline phase of the delafossite structure is the most stable among the 900 (meta)stable CZTO. Its thermodynamic stability at room temperature is also confirmed by the molecular dynamics study. Excitingly, this new phase of CZTO displays a direct band gap where the dipole-allowed transition occurs, making it a strong candidate for efficient light absorption. Furthermore, the estimation of spectroscopic limited maximum efficiency (SLME) directly demonstrates the high potential of delafossite-CZTO as a photovoltaic absorber. Our numerical results suggest that delafossite-CZTO holds another promise for future photovoltaic applications.

[a)] These authors contribute equally

[b)] Corresponding authors: smyoon@gachon.ac.kr, okjongmok@pusan.ac.kr

The efficiency of a photovoltaic cell depends heavily on the material used as an absorber. Silicon (Si) is currently the most widely used absorber due to its abundance and affordability. Though the size of its band gap is favorable for visible light absorption, it exhibits limited absorption properties in the visible spectrum owing to its indirect nature of the band gap. Accordingly, Si requires thick wafers to absorb light, leading to low power conversion efficiency (PCE) and increasing costs[1]. GaAs[2], $Cu_2InGaSe_4$ (CIGS)[3], and halide perovskites such as $CH_3NH_3PbI_3$[4] have been explored as alternative materials to overcome these limitations. GaAs and CIGS have better absorption properties than Si but are made of expensive elements such as gallium (Ga) and indium (In)[5, 6]. Halide perovskites are emerging as a promising class of photovoltaic absorbers due to their outstanding absorption properties[4]. However, they currently suffer from various instabilities associated with organic molecules. Finding new materials with superior absorption properties, high stability, and lower production prices is still a key obstacle for developing next-generation photovoltaic cells and making solar energy an alternative energy resource to fossil fuels[7, 8].

$Cu_2ZnSnS_4$ (CZTS), which possesses a zincblende-derived structure, has attracted considerable attention as a potential alternative to CIGS, primarily due to its potential to overcome the limitation of CIGS[9]. CZTS comprises earth-abundant, non-toxic, and cost-effective elements, namely zinc (Zn) and tin (Sn), and exhibit favorable properties such as a band gap size of 1.45 eV[10] and a high absorption coefficient $10^4$ cm$^{-1}$.[11] However, despite extensive research, CZTS-based solar cells have yet to surpass the efficiency of 12%[12], largely owing to issues such as phase separation and the emergence of several structural defects[13]. On the other hand, $Cu_2ZnSnO_4$ (CZTO), which consists of oxygen (O) with the same number of valence electrons as other chalcogen elements, could potentially demonstrate similar physical properties to CIGS and CZTS. Furthermore, if successfully synthesized, CZTO would offer its own advantages as oxide compounds generally exhibit high stability under various ambient environments[14]. Nonetheless, it is worth noting that, to the best of our knowledge, the thermodynamic and physical properties of a quaternary oxide CZTO have not been investigated yet in both theory and experiment.

In this paper, we explore the most stable crystal structure within the quaternary CZTO compound, known as delafossite-CZTO. Delafossite oxide belongs to a class of metal oxides where triangular A and hexagonal $BO_2$ atomic layers stack alternately, forming a unique three-dimensional structure. This stable phase is identified through the particle swam optimization (PSO) process. The delafossite-CZTO exhibits a direct band gap and dipole-allowed transitions, facilitating robust light absorption. Spectroscopic limited maximum efficiency (SLME) analysis indicates that this new phase of CZTO offers higher efficiency compared to other oxide materials. This work highlights the potential of delafossite-CZTO as an efficient and promising photovoltaic absorber with desirable properties such as high efficiency, low cost, low toxicity, and high stability.

By employing the particle swam optimization (PSO) algorithm based on *ab initio* density functional theory (DFT) calculations, we have identified the thermodynamically stable crystal structures among quaternary CZTO compounds. During the optimization process, we relaxed the volume of the crystal structure while keeping the ratio of atoms fixed (*i.e.,* Cu:Zn:Sn:O = 2:1:1:4). Among 900 crystal structures considered, those deviating by 0.7 eV/atom or less from the most stable structure are presented (denoted by a blue circle in FIG. 1(a)). The most stable structure, resembling a delafossite oxide, features alternating triangular and hexagonal layers of Cu and $Zn_{0.5}Sn_{0.5}O_2$ (denoted as delafossite-CZTO) (see FIG. 1(b)). Surprisingly, the expected stable kesterite and stannite phases, akin to CIGS and CZTS, were energetically less favorable. The total energy of kesterite and stannite structures was 200 and 204 meV/atom higher, respectively than delafossite-CZTO (as indicated by the purple and orange colors in Fig. 1(a)).

When determining the stability of a structure, it's not enough to simply rely on the strength of its cohesive energy. A structure could still be prone to spontaneous changes if there are negative frequencies in its vibrations, which would suggest structural instability. To assess this, we studied the vibrational properties of delafossite-CZTO, as shown in FIG. 2(a). The absence of negative frequencies along the high-symmetry line in the phonon band suggests that delafossite-CZTO is dynamically stable. However, while phonon dispersion can provide some insight into the structure's stability, it does not definitively tell us whether the structure might collapse at a specific temperature. To determine whether the new delafossite-CZTO structure can remain stable at room temperature (300 K), we used canonical molecular dynamics (MD) simulations using a 3×3×3 supercell. FIG. 2(c) shows how the total potential energy evolved over a 3 ps MD simulation and provides a snapshot of the final structure. Our findings confirm that delafossite-CZTO structure maintains stability at room temperature (300K).

Having checked the structural stabilities of delafossite-CZTO, we move to the electronic structure and its properties. The efficiency of a photovoltaic absorber is determined by several factors, including the bulk optical property of the absorber, the type and distribution of structural defects within the absorber, and the artifacts induced by the fabrication of the device. Among these, the intrinsic factors, i.e., the bulk optical properties, are mainly determined by the electronic structure of the absorber material. Thus, we calculated the band structure of delafossite-CZTO, which emerges as the most stable phase, using DFT calculations at the GGA-PBE and $GW_0$ levels. As shown in FIG. 2(a), delafossite-CZTO exhibits a direct band gap at the Γ point in both GGA-PBE and $GW_0$ calculations, suggesting that the direct band transition positions in lower energy than other indirect band transitions. FIG. 2(a) shows that the valence and conduction bands in delafossite-CZTO comprise Cu $d$ and Zn/Sn $p$ orbitals. The transition between the two orbitals is dipole-allowed across the direct band gap. Thus, the light is absorbed strongly when the dipole-allowed transition occurs. The size of the direct band gap is estimated to be 1.58 eV ($GW_0$), very close to that of the ideal photovoltaic absorber (~1.34 eV) according to the Shockley and Queisser (SQ) criterion[1]. These results indicate that delafossite-CZTO has a preferable electronic structure for the application of a photovoltaic absorber. To note, the gap size differs significantly by 1.32 eV for 1.58 eV ($GW_0$) and 0.26 eV (GGA-PBE). The GGA-PBE method commonly underestimates band gaps due to self-interaction errors. The $GW_0$ approach corrects this error, providing a more accurate band gap prediction.

The absorption spectrum $\varepsilon_2(\omega)$ (the imaginary part of the dielectric function) of delafossite-CZTO is estimated based on the GW calculations, as shown in FIG. 2(b). We further solve the Bethe-Salpeter equation (BSE) with quasiparticle energy bands to consider the photo-excited electron and hole interactions. We chose the twenty occupied orbitals, and six unoccupied orbitals suffice to converge $\varepsilon_2(\omega)$ up to 5.0 eV. The absorption spectra obtained from GW + BSE are compared with the non-interacting case (GW + RPA). Both GW + BSE and GW + RPA spectra do not show superior optical absorption in energy below $E_g$ (denoted as the dashed line in FIG. 2(b)), but the electron-hole interactions shift the overall absorption spectra to the lower photon energy range. The red shift of the absorption spectra above $E_g$ generally improve the efficiency of a photovoltaic absorber because the light absorption within the solar spectral range is enhanced by the redshift.

To computationally screen new absorber candidates effectively, having a descriptor that captures a photovoltaic absorber's intrinsic properties is essential. The classical descriptor is simply the band gap, which Shockley and Queisser (SQ) suggested[15]. Based on this SQ criterion, the optimal band gap is ~ 1.34 eV for the maximum solar conversion efficiency to 33.16% for a single-junction solar cell. However, it is obvious that this descriptor is insufficient because the numerous materials with similar sizes of the optimized band gap exhibit poor photovoltaic efficiency in experiments. A more sophisticated model, a widely used descriptor of the efficiency of a photovoltaic absorber, is the spectroscopic limited maximum efficiency (SLME; η)[16]. This can be estimated by DFT calculations as follows:

$$\eta = \frac{P_{max}}{P_{in}} = \frac{\max\{(j_{sc} - j_0(e^{eV/kT} - 1))V\}_V}{\int_0^\infty E I_{SUN}(E) dE}$$

Here, $j_{sc} = e \int_0^\infty a(E) I_{SUN}(E) dE$ is the short-circuit current density where $a(E) = 1 - e^{-2\alpha(E)L}$ is the photon absorptivity and $I_{SUN}(E)$ is AM1.5G solar spectrum. $j_0 = j_0^r/f_r$ is the reverse saturation current density, $j_0^r = e\pi \int_0^\infty a(E)I_{bb}(E,T)dE$ is the rates of emission and absorption through the cell surface, $I_{bb}(E,T)$ is the black-body spectrum at temperature $T$, $f_r = e^{(E_g - E_g^{da})/kT}$ is the fraction of the radiative recombination current, and $E_g$ and $E_g^{da}$ denote the band gap and direct-allowed band gap. Accordingly, the property-related inputs are the absorption coefficient $\alpha(E)$, the fraction of the radiative recombination current $f_r$, the thickness $L$, and the temperature $T$. Note that extrinsic effects such as structural defects and fabrication-induced artifacts are not considered in the SLME.

The SLME of delafossite-CZTO is calculated as a function of thickness to evaluate the potential efficiency of this new material, as shown in FIG. 3(b). For the thickness $L$ = 2.0 μm and the temperature $T$ = 300K, the SLME of delafossite-CZTO is about 28.2 %. The SLME of CZTO is relatively high compared to other oxide materials proposed as promising photovoltaic absorbers (see Table I)[17-23]. The SLME in Table I is evaluated for the thickness $L$ = 2.0 μm and temperature $T$ = 300K; the SLME is strongly dependent on the thickness and the temperature of the photovoltaic absorber. The estimation of the SLME directly suggests that delafossite-CZTO could be an efficient photovoltaic absorber. Here, it is worth recalling that CZTO has several additional advantages for real applications, such as element abundance, low cost, low toxicity, and high stability under an ambient environment.

To synthesize the promising quaternary compound CZTO, we have tried high-temperature sintering up to $T$ = 1673 K and under ambient conditions using an alumina crucible. FIG. S1 shows the X-ray diffraction (XRD) $2\theta$ scan of the synthesized specimen. The XRD pattern is unveiled not to be analyzed with theoretically predicted CZTO structures. Rather, the XRD pattern can be interpreted with four stable binary and ternary compounds: $Zn_2SnO_4$, $SnO_2$, $Cu_2O$, and CuO. The XRD peak intensities of $Cu_2O$ and CuO are relatively weaker than those of $Zn_2SnO_4$ and $SnO_2$, which might be attributed to the volatile nature of Cu atoms and the reaction between $Cu_xO$ and the alumina crucible[24]. This result indicates that the precise control of stoichiometry (particularly the amount of Cu atoms) is crucial for synthesizing CZTO. Still, there are various alternative methods to try synthesizing CZTO compounds; *for example,* quartz sealing, pressure-controlled heat treatment, and rapid liquid phase synthesis through arc discharge, as well as thin-film synthesis via physical vapor deposition methods like sputtering and pulsed laser deposition (PLD). The experimental validation of our work remains for future work.

Finding efficient and cost-effective materials for photovoltaic absorbers is crucial for advancing solar energy technology and achieving sustainable energy solutions. Our study explored CZTO compounds as promising alternatives to conventional absorber materials, employing *ab initio* DFT calculations and PSO algorithm. We identified the most stable crystal structure, known as delafossite structure, which exhibited a direct band gap conducive to effective light absorption. SLME analysis underscored CZTO's high efficiency potential. If the discovered phase of CZTO is synthesized, it will present a compelling option for next-generation photovoltaic cells, with the material's abundance, low cost, low toxicity, and potential stability under operating conditions. Continued research into CZTO and other innovative materials will contribute to developing efficient and affordable solar energy systems, driving the global shift towards renewable energy and reducing dependence on fossil fuels.

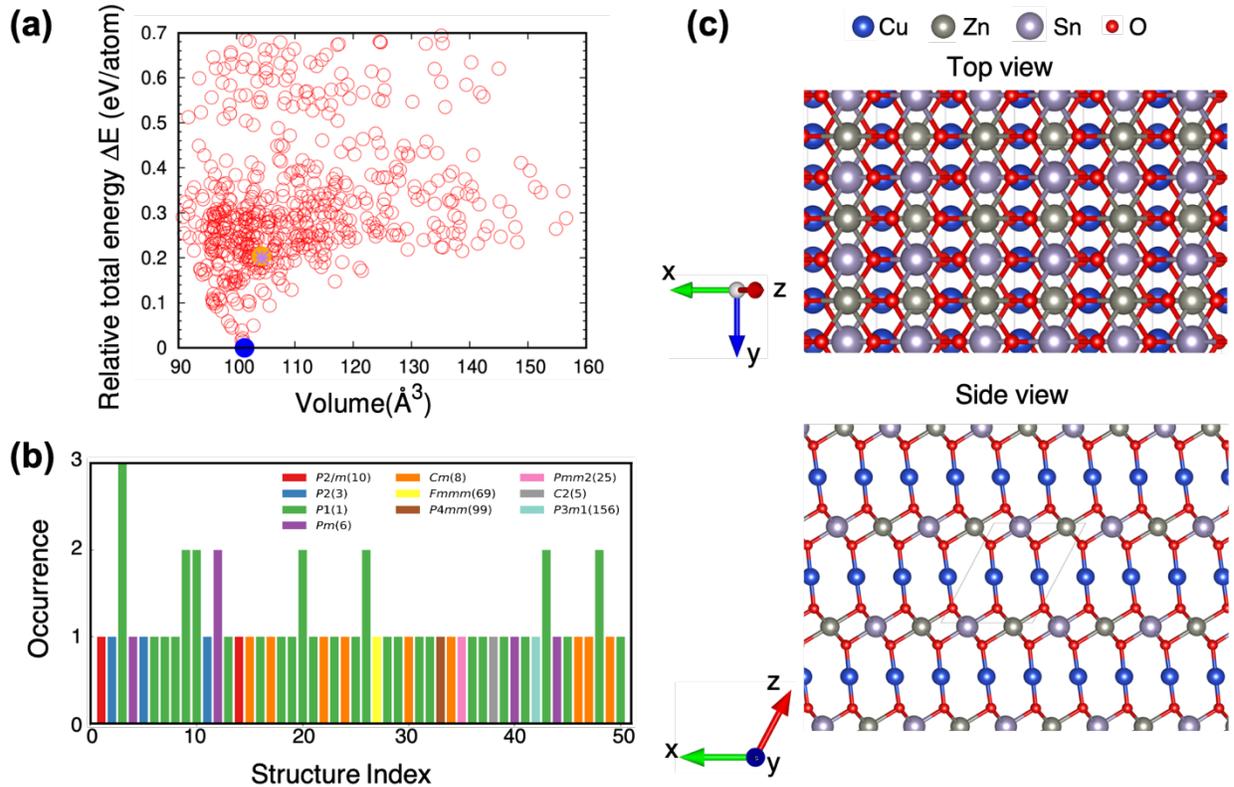

FIG. 1. Computational search for thermodynamically stable $Cu_2ZrSnO_4$ (CZTO) (a) Relative total energy $\Delta E$ with respect to the most stable structure as a function of volume for CZTO. The stable structures are found by using the particle swarm optimization (PSO) algorithm. The most stable structure is color-coded blue. (b) The low-energy structures of CZTO are found by the PSO algorithm. The crystal structures are color-coded and ordered according to the energy hierarchy, i.e., the first one is the most stable structure. (c) Top and side views of the most stable structure within our materials search in CZTO: delafossite structure. The black solid line shows a unit cell of the structure.

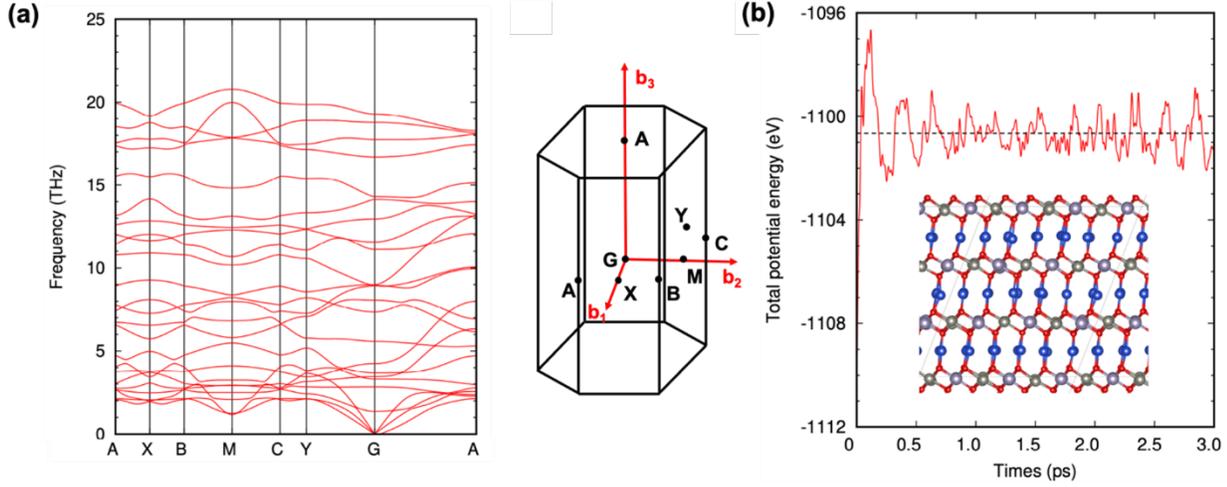

FIG. 2. Structural stability of delafossite-CZTO (a) Phonon band structures along the high-symmetry points (b) Total potential energy as a function of time during canonical MD simulations at room temperature with the final structural at the end of the simulation time of 3 ps (inset). The black-dashed line represents the potential average from the MD simulation at room temperature.

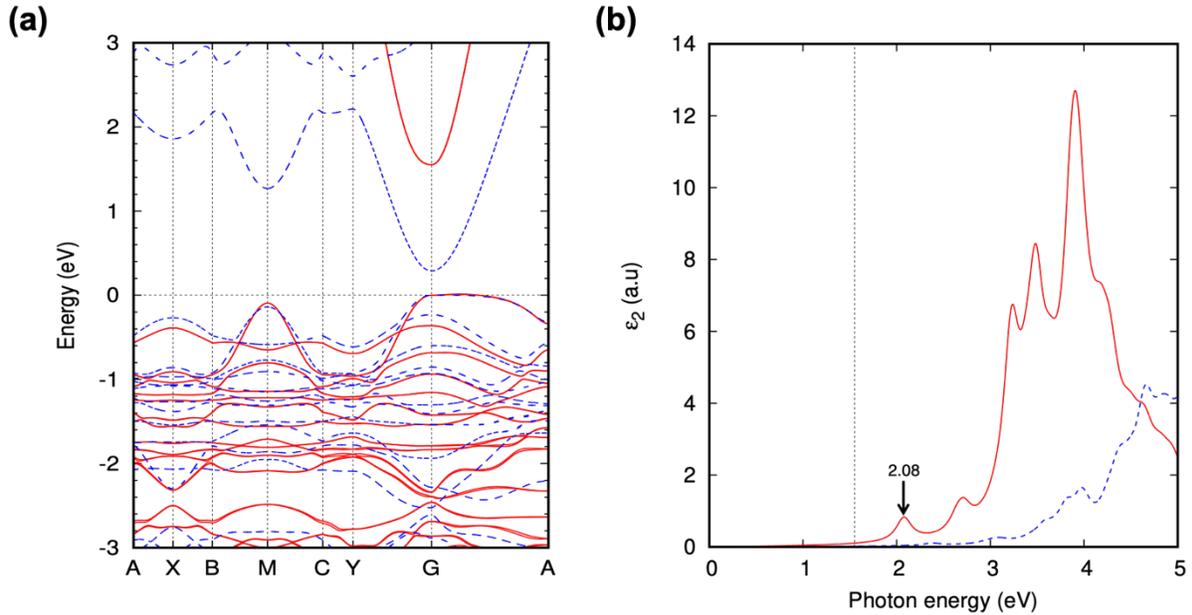

FIG. 3. Electronic structure of delafossite-CZTO (a) Band structure of delafossite-CZTO. Blue-dashed and red-solid lines represent energy bands calculated from DFT-PBE and $GW_0$, respectively. The valence band maximum is set to zero. (b) Absorption spectra $\varepsilon_2$ from GW + BSE (red solid line) and GW + RPA (blue dashed line). The dotted gray lines indicate photon energy equal to the quasiparticle energy gap.

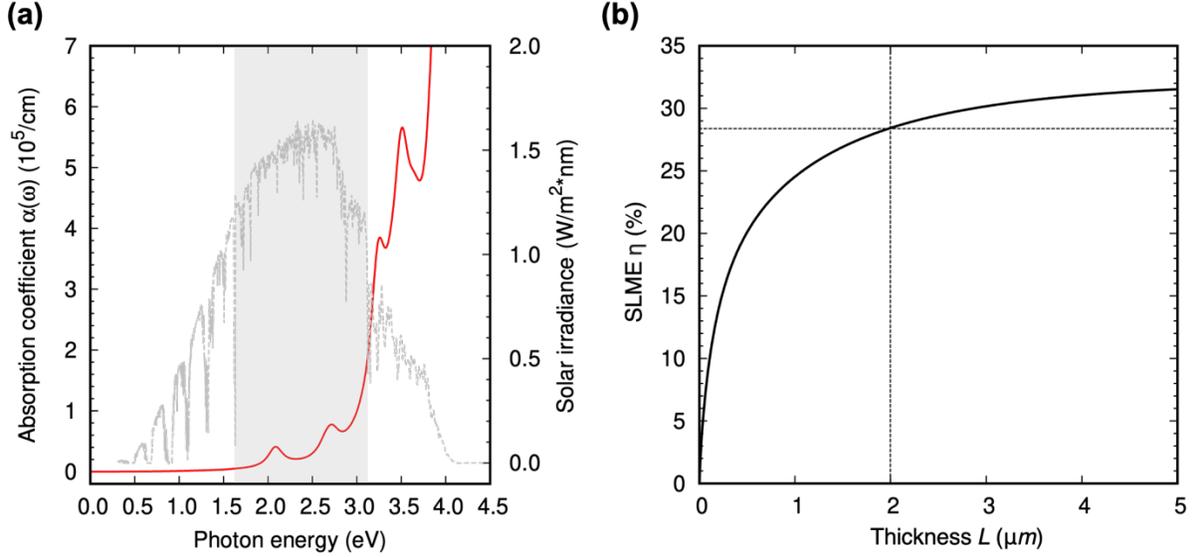

FIG. 4. Absorption properties and figure-of-merit of delafossite-CZTO (a) Absorption coefficient from GW + BSE for bulk $Cu_2ZnSnO_4$. The gray box indicates the energy range of visible light. The AM1.5 is shown with gray lines. (b) The SLME ($\eta$) as a function of thickness L at 300K.

Table I. Spectroscopic limited maximum efficiency (SLME) of oxide compounds proposed as promising photovoltaic absorbers. The SLME are evaluated for the thickness $L = 2.0$ μm and temperature $T = 300K$. The SLME of CdTe is also included for reference[17-23].

|  | SLME (%) |  | SLME (%) |
|---|---|---|---|
| $Cu_2ZnSnO_2$ | 28.2 | $Cu_2O$ | 0.5 |
| $CuGaO_2$ | 32.6 | $Cu_2O$:Zn | 8.0 |
| $CuInO_2$ | 31.9 | $Ba_2SnNbO_6$ | 26.5 |
| SnO | 5.3 | $Na_2Tl_{0.25}Bi_{0.75}O_6$ | 15.5 |
| 2D-SnO | 13.4 | $SrBaVBiO_6$ | 16.8 |
| $Sn_{0.75}Zn_{0.25}O$ | 22.2 | CdTe (ref) | 30.1 |

## METHODS

*Crystal prediction calculation*

To identify energetically stable structures of CZTO, we perform crystal prediction calculations using the particle swarm optimization (PSO) algorithm implemented in Crystal Structure Analysis by Particle Swarm Optimization (CALYPSO)[25]. At the beginning of the simulation, we start with 50 random structures within 3D space groups, and the structural evolution proceeds up to 6 generations based on the PSO scheme for the three different volumes. A total of 900 structures were generated without any symmetry constraint. Out of 900 initial structures, we obtained 503 optimized structures. The total energies of the configurations are calculated using the all-electron full-potential FHI-aims code[26-28] with the Perdew-Burke-Ernzerhof (PBE) exchange-correlation functional[29]. The tight numerical settings and 4 × 4 × 4 k-point grids are used in these calculations. All structures are fully optimized using the Broyden-Fletcher-Goldfarb-Shanno (BFGS)[30] algorithm, with a maximum force component below $10^{-3}$ eV/Å. After that, we also checked the structure optimization with more dense k-point grids 8 × 8 × 8 using the Vienna Ab-initio Simulation Package (VASP)[31, 32].

*Ab initio density functional theory (DFT) calculations*

In our present investigation of the electronic and optical properties of CZTO, we employ Density Functional Theory (DFT) with the projector-augmented wave (PAW) method[33], as implemented in the Vienna Ab-initio Simulation Package (VASP)[31, 32]. The exchange and correlation functional used the generalized gradient approximation (GGA) in the Perdew-Burke-Ernzerhof (PBE) form[29]. A cutoff of 420 eV was set for the plane-wave expansion of the wave function. Our GW calculations are of the $GW_0$[34, 35] type and include self-consistency with five iterations. To ensure convergence of the relevant quantities, we considered a total of 450 bands. For $GW_0$ calculations, the Brillouin zone (BZ) sampling was conducted with a 5 × 5 × 5 grid. We also solved the Bethe-Salpeter Equation (BSE)[36-41] using quasiparticle energy bands to evaluate the optical absorption spectrum, $\varepsilon_2(\omega)$ (the imaginary part of the dielectric function), and exciton energy levels of CZTO. We considered the top twenty occupied and six unoccupied orbitals to achieve convergence of $\varepsilon_2(\omega)$ up to 5.0 eV. The absorption spectra obtained with electron-hole interaction (GW +BSE) are compared with those obtained in the non-interacting case ($GW_0$)

*Synthesis and Characterization*

In this study, the bulk samples were prepared via the solid-state reaction method. To begin, we prepared powders of $Cu_2O$, ZnO, and $SnO_2$ in the molar ratio of 1:1:1, ensuring an error level of ±0.5 mg during weighing. These powders were mixed thoroughly and pressurized at 10 MPa for 5 min to form a pellet. The prepared pellets were reacted in the atmosphere using a box furnace. The heat treatment process involved raising the temperature from 295 to 1475 K at a rate of 200 K/h and holding it at 1475 K for 16 hours. The crystal structure of the bulk sample was examined by X-ray diffraction (XRD) with Cu radiation (X'pert3 X-ray Diffractometer). Diffraction patterns were collected at room temperature.

**ACKNOWLEDGMENT**

This work was supported by the U.S. Department of Energy (DOE), Office of Science, National Quantum Information Science Research Centers, Quantum Science Center (S.-H. K.) and by the U.S. Department of Energy, Office of Science, Office of Basic Energy Sciences, Materials Sciences and Engineering Division (M. Y.) and this research used resources of the Oak Ridge Leadership Computing Facility at the Oak Ridge National Laboratory, which is supported by the Office of Science of the U.S. Department of Energy under Contract No. DE-AC05-00OR22725 and resources of the National Energy Research Scientific Computing Center, a DOE Office of Science User Facility supported by the Office of Science of the U.S. Department of Energy under Contract No. DE-AC02-05CH11231 using NERSC award BES-ERCAP0024568. This research was supported by the 2023 BK21 FOUR Program of Pusan National University (J. M. O.). This work was supported by the National Research Foundation of Korea (NRF) grant funded by the Korea government (MSIT) (No. RS-2023-00210295) (J. M. O.). This work was supported by the National Research Foundation of Korea (NRF) (NRF-2022R1F1A1072330) (S. Y.). This research was supported by the Gachon University research fund (GCU-2021-1034) (S. Y.).